\documentclass[aps,prl,reprint]{revtex4-1}
\usepackage[mathlines]{lineno}
\usepackage{lipsum}
\usepackage{blindtext}
\usepackage[export]{adjustbox}
\usepackage{graphicx,epsfig,epstopdf}
\usepackage{amsmath,amssymb}
\usepackage{esvect}
\usepackage{mathtools}
\usepackage{color}
\usepackage{subfigure}
\usepackage{array}
\usepackage{tabularx}
\usepackage{multirow}
\usepackage{booktabs}
\usepackage{float}
\newcolumntype{L}[1]{>{\raggedright\arraybackslash}p{#1}}
\newcolumntype{C}[1]{>{\centering\arraybackslash}p{#1}}
\newcolumntype{M}[1]{>{\centering\arraybackslash}m{#1}}
\allowdisplaybreaks

\def\_#1{{\bf #1}}
\def\.{\cdot}

\def\Re{{\rm Re\mit}}
\def\Im{{\rm Im\mit}}

\begin{document}

\preprint{}

\title{Perfect Generation of Angular Momentum with Cylindrical  Bianisotropic Metasurfaces}

	\author{Junfei Li$^{1,\star}$, Ana~D\'{i}az-Rubio$^{2,\star}$, Chen Shen$^{1}$, Zhetao Jia$^{1}$, Sergei~Tretyakov$^{2}$,   and Steven Cummer$^1$}
	
	\affiliation{$^1$Department of Electrical and Computer Engineering, Duke University, Durham, North Carolina 27708, USA\\
		$^2$Department of Electronics and Nanoengineering, Aalto University, P.~O.~Box~15500, FI-00076 Aalto, Finland\\
        $^\star$ J.L. and A.D. contributed equally to this work
        }

\date{\today}

\begin{abstract}
Recent advances in metasurfaces have shown the importance of controlling the bianisotropic response of the constituent meta-atoms for maximum efficiency wavefront transformation. Under the paradigm of a bianisotropic metasurface, full control of the local scattering properties is allowed opening new design avenues for creating reciprocal metasurfaces. Despite recent advances in the perfect transformation of both electromagnetic and acoustic plane waves, the importance of bianisotropic metasurfaces for transforming cylindrical waves is still unexplored. Motivated by the possibility of arbitrarily controlling the angular momentum of cylindrical waves, we develop a design methodology for a bianisotropic cylindrical metasurface that enables perfect transformation of cylindrical waves. This formalism is applied to the acoustic scenario and the first experimental demonstration of perfect angular momentum transformation is shown. 

\end{abstract}

\maketitle

\section{Introduction}

Metamaterials have been serving as a primary approach to fully control the behavior of electromagnetic waves, acoustic waves and elastic waves in recent years \cite{engheta2006metamaterials,cummer2016controlling}, and is at present a highly active research area. Metasurfaces, as the 2D version of metamaterials, have opened up unprecedented possibilities for controlling waves at will, offering a solution of molding wave propagation within a planar geometry \cite{chen2016review,glybovski2016metasurfaces}. By engineering the local phase shift in the unit cells, various functionalities have been achieved by metasurfaces, such as focusing \cite{pors2013broadband}, wave redirection and retro-reflection \cite{yu2011light,kildishev2013planar,arbabi2017planar}, enhanced absorption \cite{yao2014electrically}, cloaking \cite{ni2015ultrathin}, and holographic rendering \cite{ni2013metasurface,zheng2015metasurface}, to name a few. However, the efficiency of phase-gradient metasurfaces is fundamentally limited by the impedance mismatch between incident field and reflected/transmitted field, so that some of the energy is scattered into unwanted higher order diffracted modes, which hinders the applicability in various scenarios.

Recent advances have demonstrated that for electromagnetic and acoustic waves, full control of refraction or reflection can be achieved by carefully controlling the bianisotropy \cite{wong2016reflectionless,asadchy2016perfect,epstein2016synthesis,estakhri2016wave,diaz2017from,diaz2017acoustic,li2018systematic}, also called Willis coupling in elastodynamics \cite{muhlestein2017experimental}, in the unit cells. By tuning both transmitted and reflected phase profiles, one can not only control the microscopic phase profile along the metasurface but also achieve the overall macroscopic impedance match between the incident and scattered fields. Such metasurfaces, called bianisotropic gradient metasurfaces, serve as the second generation of metasurfaces for wavefront manipulation \cite{asadchy2018bianisotropic}. In recent studies of wave deflection with both electromagnetic and acoustic bianisotropic gradient metasurfaces, it has been shown that the transmission efficiency can be significantly improved, especially for large deflection angles. Also, it has been demonstrated that bianisotropic gradient metasurfaces offer scattering-free wave manipulation even with a relatively coarse piecewise approximation of the required impedance matrix profile \cite{li2018systematic}, which provides advantages in fabrication. However, the concept of bianisotropic metasurfaces and systematic design for scattering-free manipulation have been less explored in other topologies. Cylindrical topologies are among the most commonly used structures in electromagnetics, acoustics, and elastodynamics. However, the concept and benefits of bianisotropic metasurfaces have not been extended to this field yet. 

In analogy to anomalous refraction for flat metasurfaces, one of the possibilities offered by cylindrical metasurfaces is the transformation between different cylindrical waves. This transformation was achieved by locally controlling the phase profile along the surface and contribute to the generation of source illusion \cite{liu2017source}.
Generation of angular-momentum waves using a  single layer of generalized Snell´s law (GSL) based metasurface will not only introduce a large impedance mismatch but will also require a fine discretization of the surface which is not easily achievable by conventional cell architectures. Therefore, generation of wave fields with a large angular momentum still remains challenging. The successful realization of scattering-free bianisotropic planar metasurfaces suggests that scattering-free cylindrical metasurfaces might be possible.

Source illusion is just an example of many possibilities offered by metasurfaces capable of controlling angular momentum.
Recent research has also demonstrated the manipulation of beams for particle trapping \cite{baresch2016observation,he1995direct} and boosting communication efficiency \cite{wang2017terabit,shi2017high} with acoustic angular momentum. 
Passive generation of wave fields with non-zero angular momentum is typically implemented by leaky wave antennas or metasurfaces based on generalized Snell's law (GSL) \cite{naify2016generation,jiang2016convert} for acoustic waves and inhomogeneous anisotropic media \cite{marruci2006optical}, spatial light modulator or spiral phase plates \cite{Yao2011Orbital,schemmel2014modular} for electromagnetic waves.
However, the recent advances in metasurfaces for perfect wavefront manipulation show that if only the transmission phase profile is controlled, parasitic scattering inevitably appears, which reduces the efficiency, or even cause the failure to the structures, especially for large angular momentum.

In this paper, we present the first theoretical study, simulation, and experimental demonstration of perfect angular momentum generation by cylindrical bianisotropic metasurfaces. 
In particular, the work is focused on metasurfaces for the manipulation of cylindrical acoustic waves (the reader is referred to the Supplementary Material for the electromagnetic counterpart). First, we theoretically analyze the generation of angular momentum showing that bianisotropic response is required for perfect performance. Next, we propose a possible realization of the required impedance matrix profile. We take an example of the transformation between a point source (zero angular momentum) and a field with an angular momentum $n=12$ and confirm in simulations that the desired field distribution is indeed created without any reflection and scattering. Finally, a realization in acoustics is verified by experiments.

\section{Results}

\textbf{Theoretical formulation.}
For acoustic waves, the 2D wave equation in the cylindrical coordinates is written as
$\nabla^2p=\frac{1}{r}\frac{\partial}{\partial r}\left(\frac{\partial p}{\partial r}\right)+\frac{1}{r^2}\frac{\partial^2p}{\partial \varphi^2}=\frac{1}{c_0^2}\frac{\partial^2p}{\partial t^2}$, where $p$ is the acoustic pressure and $c_0$ is the sound speed.
Just like plane waves in Cartesian coordinates, Bessel-like spinning waves with different angular momentum serve as the bases in cylindrical coordinates. 
In the general case, the solution to this equation can be written as $p=\sum_{n} {\left[a_nH_n^{(1)}(kr)+b_n H_n^{(2)}(kr)\right] e^{\mathrm{j}n\varphi}e^{\mathrm{j}\omega t} }$, where $H_n^{(1)}$ denotes the Hankel function of the first kind (waves converging to the center) and $H_n^{(2)}$ denotes the Hankel function of the second kind (waves diverging from the center), index $n$ represents the angular momentum, $a_{n}$ and $b_{n}$ are the amplitudes of the waves, and $k=\omega/c_0$ is the wavenumber at the frequency of interest. 

\begin{figure}
\includegraphics[width=0.85\linewidth]{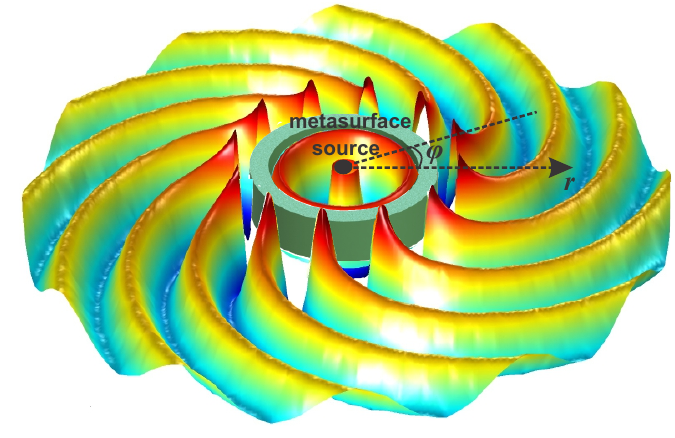}
\caption{Illustration of the desired performance of a metasurface to convert the inner field to a prescribed outer field without parasitic scattering}
\label{fig:Fig1}\end{figure}

In this section we will discuss the theoretical requirements for a metasurface to produce perfect transformation between cylindrical waves with different angular momenta, i.e. with different spinning characteristics, as it is shown in Fig.~\ref{fig:Fig1}. The derivation of the solution will be presented considering acoustic waves, however, a similar formulation can be used for electromagnetic waves (see Supplementary Note 1).

The formulation of the problem starts with the definition of the fields inside and outside the volume bounded by the metasurface. In what follows, the time-harmonic dependency $e^{\mathrm{j}\omega t}$ will be omitted. Let us consider the field in Medium~I (inside the metasurface) as a divergent wave with the angular momentum $n_1$ that can be expressed as $p^{\rm I}=p_0H_{n_1}^{(2)}(kr)e^{\mathrm{j}{n_1}\varphi}$, where $p_0$ is the wave amplitude. It is important to mention that we only consider a divergent wave inside the metasurface because the objective of the metasurface is to perfectly transform the incident cylindrical wave without reflections. The velocity vector can be calculated from the pressure field ($\vv{v}=\frac{\mathrm{j}}{\omega \rho} \nabla p$) as $\vv{v}^{\rm I}= \frac{p_0}{Z_0}\left[\mathrm{j}\partial_rH_{n_1}^{(2)}(kr)\hat{\rho}-\frac{n_1}{kr}H_{n_1}^{(2)}(kr)\hat{\varphi}\right]e^{\mathrm{j}n_1\varphi}$, where $Z_0=\sqrt[]{c_0\rho}$ is the characteristic impedance of air and $\partial_r$ represents the partial derivative with respect to $r$. Following the same approach, we can define the field in Medium~II (outside the volume bounded by the metasurface) as a divergent wave with the angular momentum $n_2$ as $p^{\rm II}=p_{\rm t}H_{n_2}^{(2)}(kr)e^{\mathrm{j}{n_2}\varphi}$ with $p_{\rm t}$ being the amplitude of the transmitted wave. The velocity vector can be expressed as $\vv{v}^{\rm II}= \frac{p_{\rm t}}{Z_0}\left[\mathrm{j}\partial_rH_{n_2}^{(2)}(kr)\hat{\rho}-\frac{n_2}{kr}H_{n_2}^{(2)}(kr)\hat{\varphi}\right]e^{\mathrm{j}n_2\varphi}$. 

We assume that the metasurface is a cylindrical tube whose axis is located at the origin, with the inner radius $r_1$ and the outer radius $r_2$.  For lossless and scattering-free metasurfaces, the energy conservation condition shall be met.  Denoting the time-averaged intensity vector as $\vv{I}=\frac{1}{2}\mathrm{Re}\left\lbrace p \vv{v}^{*}\right\rbrace= I_r \hat{\rho} + I_\varphi \hat{\varphi}$, this condition can be expressed in terms of the radial components of this vector at the two sides of the metasurface:
\begin{eqnarray}
& I_r^{\rm I}=\frac{p_0^2}{2Z_0}\left[J_{n_1}(kr)\partial_rY_{n_1}(kr)-Y_{n_1}(kr)\partial_rJ_{n_1}(kr)\right]|_{r_1}\\
& I_r^{\rm II}=\frac{p_{\rm t}^2}{2Z_0}\left[J_{n_2}(kr)\partial_rY_{n_2}(kr)-Y_{n_2}(kr)\partial_rJ_{n_2}(kr)\right]|_{r_2},
\end{eqnarray}
where $J_\alpha$ and $Y_\alpha$ represent the Bessel functions of the first and second kind, respectively. These expressions can be simplified as $I_r^{\rm I}=\frac{p_0^2}{\pi Z_0}\frac{1}{r_1}$ and $I_r^{\rm II}=\frac{p_{\rm t}^2}{\pi Z_0}\frac{1}{r_2}$.

To ensure that all the energy of the incident wave is carried away by the transmitted spinning wave, the normal component of the intensity vector crossing a line element of the inner radius, $S_1=r_1 {\rm d}\varphi$,  has to be equal to the one crossing the corresponding line element in the other radius, $S_2=r_2 {\rm d}\varphi$. This condition can be written as $ I_r^{\rm I} S_1= I_r^{\rm II} S_2$ and yields that the  $p_{\rm t}=p_0$. 
If we define the macroscopic transmission coefficient as
\begin{equation}
 T=\frac{p^{\rm II}(r_2)}{p^{\rm I}(r1)}=\frac{H_{n_2}^{(2)}(kr_2)}{H_{n_1}^{(2)}(kr_1)}e^{j(n_2-n_1)\varphi}, 
\end{equation}
it is possible to see that the magnitude of the macroscopic transmission coefficient can be greater (smaller) than unity if $n_2$ is greater (smaller) than $n_1$, respectively. 
It is noted here that this condition is analogous to the plane-wave case described in \cite{li2018systematic,diaz2017acoustic}.

The next step towards the realization of perfect transformation between cylindrical waves is to determine the required boundary conditions at both sides of metasurface.
At the inner and outer boundaries of the metasurface, for each specific circumferential position, the impedance matrix which models the metasurface is defined as
\begin{equation}
\begin{bmatrix}
p^{\rm I}(r_1,\phi)    \\
p^{\rm II}(r_2,\phi)
\end{bmatrix}=
\begin{bmatrix}
Z_{11}     & Z_{12}  \\
Z_{21}     & Z_{22}
\end{bmatrix}\begin{bmatrix}
\;\;\; S_1\hat{n}\cdot\vv{v}^{\rm I}(r_1,\phi)    \\
-S_2\hat{n}\cdot\vv{v}^{\rm II}(r_2,\phi)
\end{bmatrix},
\label{eq:condition}
\end{equation}
where $\hat{n}$ is the unit normal vector to the metasurface. 
In the most general linear, time-invariant, lossless, and reciprocal case, the impedance matrix is symmetric, $Z_{12}=Z_{21}$,  and all its components are purely imaginary, $Z_{ij}=jX_{ij}$ \cite{li2018systematic}. 

For compactness, we denote 
$C_{n_1}=H_{n_1}^{(2)}(kr_1)e^{\mathrm{j}{n_1}\phi}$,
$C_{n_2}=H_{n_2}^{(2)}(kr_2)e^{\mathrm{j}{n_2}\phi}$,
$C_{n_1}^\prime=\frac{1}{2}[H_{{n_1}-1}^{(2)}(kr_1)-H_{{n_1}+1}^{(2)}(kr_1)]e^{\mathrm{j}{n_1}\phi}$ and
$C_{n_2}^\prime=\frac{1}{2}[H_{{n_2}-1}^{(2)}(kr_2)-H_{{n_2}+1}^{(2)}(kr_2)]e^{\mathrm{j}{n_2}\phi}$,
and re-write Eq.~(\ref{eq:condition}) in form of a system of two linear equations:
\begin{equation}
\left\{ \begin{array}{lr}  
  Z_0C_{n_1}=-S_1 X_{11} C^\prime_{n_1} +S_2 X_{12} C^\prime_{n_2} &  \\  
  Z_0C_{n_2}=-S_1 X_{12} C^\prime_{n_1} +S_2 X_{22} C^\prime_{n_2} & 
             \end{array}  
\right.
\end{equation}
After some algebra, the components of the impedance matrix can thus be calculated:
\begin{eqnarray}
X_{11}=\frac{Z_0}{S_1}\frac{\Im(C_{n_1}) \Re(C_{n_2}^\prime) -\Re(C_{n_1})\Im(C_{n_2}^\prime)}{\Im(C_{n_2}^\prime)\mathrm{Re}(C_{n_1}^\prime)-\Re(C_{n_2}^\prime)\Im(C_{n_1}^\prime)}\label{eq:X11}\\
X_{22}=\frac{Z_0}{S_2}\frac{\mathrm{Im}(C_{n_2})\mathrm{Re}(C_{n_1}^\prime)-\mathrm{Re}(C_{n_2})\mathrm{Im}(C_{n_1}^\prime)}{\mathrm{Im}(C_{n_2}^\prime)\mathrm{Re}(C_{n_1}^\prime)-\mathrm{Re}(C_{n_2}^\prime)\mathrm{Im}(C_{n_1}^\prime)}\label{eq:X22} \\
X_{12}=-\frac{Z_0}{S_1}\frac{\mathrm{Im}(C_{n_2}^\prime)\mathrm{Re}(C_{n_2})-\mathrm{Re}(C_{n_2}^\prime)\mathrm{Im}(C_{n_2})}{\mathrm{Im}(C_{n_2}^\prime)\mathrm{Re}(C_{n_1}^\prime)-\mathrm{Re}(C_{n_2}^\prime)\mathrm{Im}(C_{n_1}^\prime)}.\label{eq:X12}
\end{eqnarray}

For simplicity in the derivations, and to provide another view point for the requirements, the required properties of the metasurface can also be expressed in terms of the transfer matrix, which is defined by
\begin{equation}
\begin{bmatrix}
p^{\rm I}(r_1,\phi)    \\
S_1\hat{n}\cdot\vv{v}^{\rm I}(r_1,\phi)
\end{bmatrix}=
\begin{bmatrix}
M_{11}     & M_{12}  \\
M_{21}     & M_{22}
\end{bmatrix}\begin{bmatrix}
  p^{\rm II}(r_2,\phi)   \\
S_2\hat{n}\cdot\vv{v}^{\rm II}(r_2,\phi)
\end{bmatrix}
\end{equation}
Conversion from the impedance matrix to the transfer matrix is given by
\begin{equation}
M=
\begin{bmatrix}
\frac{Z_{11}}{Z_{21}}     & \frac{Z_{11}Z_{22}-Z_{21}Z_{12}}{Z_{21}}  \\
\frac{1}{Z_{21}}     & \frac{Z_{22}}{Z_{21}}
\end{bmatrix}\label{eq:M}
\end{equation}
which indicates that $M_{11}$ and $M_{22}$ are real, while $M_{12}$ and $M_{21}$ are imaginary. Explicit solution for the transfer matrix can thus be written:
\begin{eqnarray}
M_{11}=\frac{\mathrm{Im}(C_{n_2}^\prime)\mathrm{Re}(C_{n_1})-\mathrm{Re}(C_{n_2}^\prime)\mathrm{Im}(C_{n_1})}{\mathrm{Im}(C_{n_2}^\prime)\mathrm{Re}(C_{n_2})-\mathrm{Re}(C_{n_2}^\prime)\mathrm{Im}(C_{n_2})}\\
M_{22}=\frac{-S_1}{S_2}\frac{\mathrm{Im}(C_{n_2})\mathrm{Re}(C_{n_1}^\prime)-\mathrm{Re}(C_{n_2})\mathrm{Im}(C_{n_1}^\prime)}{\mathrm{Im}(C_{n_2}^\prime)\mathrm{Re}(C_{n_2})-\mathrm{Re}(C_{n_2}^\prime)\mathrm{Im}(C_{n_2})}\\
M_{12}=\frac{\mathrm{j}Z_0}{S_2}\frac{\mathrm{Im}(C_{n_2})\mathrm{Re}(C_{n_1})-\mathrm{Re}(C_{n_2})\mathrm{Im}(C_{n_1})}{\mathrm{Im}(C_{n_2}^\prime)\mathrm{Re}(C_{n_2})-\mathrm{Re}(C_{n_2}^\prime)\mathrm{Im}(C_{n_2})}\\
M_{21}=\frac{\mathrm{j}S_1}{Z_0}\frac{\mathrm{Im}(C_{n_2}^\prime)\mathrm{Re}(C_{n_1}^\prime)-\mathrm{Re}(C_{n_2}^\prime)\mathrm{Im}(C_{n_1}^\prime)}{\mathrm{Im}(C_{n_2}^\prime)\mathrm{Re}(C_{n_2})-\mathrm{Re}(C_{n_2}^\prime)\mathrm{Im}(C_{n_2})}.
\end{eqnarray}
It is easy to check that this matrix corresponds to a reciprocal and lossless system. 

Note that as long as $|n_1|\neq |n_2|$, we will always have $M_{11}\neq M_{22}$, which leads to $Z_{11}\neq Z_{22}$. This asymmetry is analogous to the plane-wave case in the Cartesian coordinates, meaning that controlling only the transmission phase along the metasurface is not enough for perfect engineering the power flow. Instead, a bianisotropic metasurface with precisely controlled asymmetric response is required. 



\textbf{Design and Realization.}
For the actual implementation of the metasurface described in the previous section, there are several different possible approaches. 
First, one can always use three membranes separated by a fixed distance. By controlling the thickness and in-plane tension of the membranes, one can, in principle, control the impedances to satisfy Eqs.~(\ref{eq:X11})-(\ref{eq:X12}). However, the surface tension, uniformity and durability for the membranes are extremely hard to control, and it is questionable whether such configuration can be practically realized. More details about this approach are given in the Supplementary Materials.

An alternative approach based on a straight channel with four resonators was proposed for flat surfaces \cite{li2018systematic}. The design provides enough degrees of freedom for full control over the bianisotropic response while reducing the loss induced by resonances. 
Here, we propose the four-resonator design in cylindrical coordinates for full control over the bianisotropic response of the unit cells. An example cell is shown in Fig.~\ref{fig:Fig3}. In this structure: the width and height of the neck, $h_{\mathrm{neck}}$  and $w_{\mathrm{neck}}$, are fixed for the four resonators; the width of the cavities $w_{cav}$ is also fixed; the sector angle of the wedge-shaped channel $\theta_c$ and the height of the resonators $w_{\rm a}$, $w_{\rm b}$, $w_{\rm c}$, and $w_{\rm d}$ can be varied to control the overall impedance response; and the wall thickness of the unit cell is fixed and will be defined by the fabrication limitations. The walls between adjacent cells are assumed to be hard so that the wave does not propagate along the orthogonal direction inside the metasurface. Therefore, all the cells in the bianisotropic metasurfaces can be designed individually.

\begin{figure}
\includegraphics[width=0.95\linewidth]{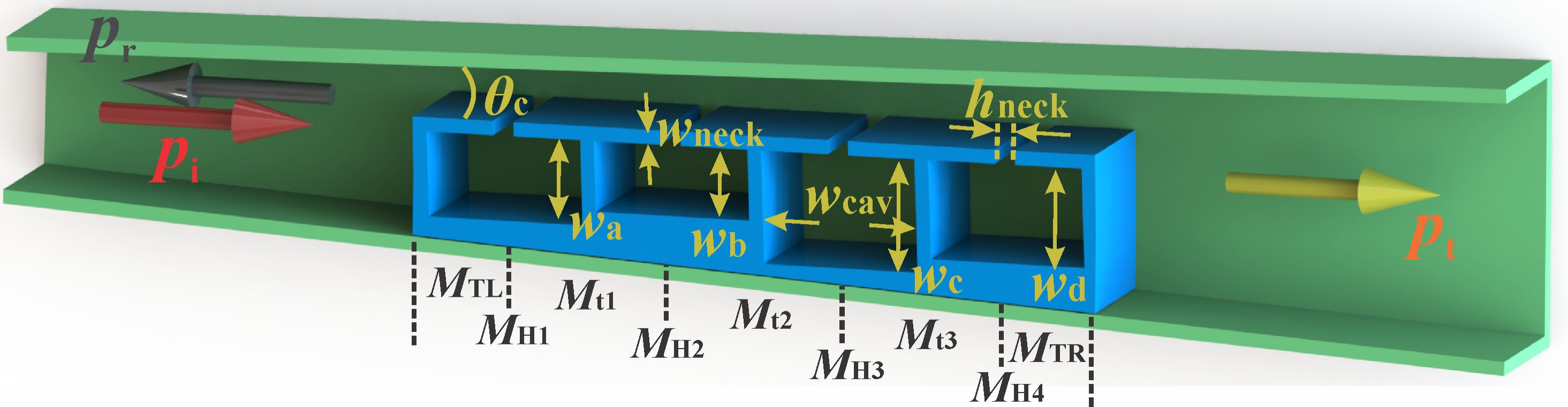}
\caption{Unit cell consisting four resonators for the realization of the impedance matrix in cylindrical coordinates.} 
\label{fig:Fig2}\end{figure}

The transfer matrix of the proposed meta-atom topology can be calculated as 
\begin{equation}
M=M_{TL}M_{H1}M_{T1}M_{H2}M_{T2}M_{H3}M_{T3}M_{H4}M_{TR}
\end{equation}
with $M_{TL}$, $M_{TR}$, and $M_{T1,2,3}$ being the transfer functions of transmission lines at the entrance, exit, and between adjacent resonators, as is shown in Fig.~\ref{fig:Fig2}.
\begin{equation}
M_{Hi}=
\begin{bmatrix}
1     & 0  \\
1/Z_{Hi}     & 1
\end{bmatrix}, \qquad i=1, 2, 3,
\end{equation}
where $Z_{\it{Hi}}$ are the acoustic impedances for each shunted resonator. The detailed derivation of $Z_{\rm {H}i}$ is given in \cite{li2016theory}, and the result is  directly given here for brevity: $Z_{\rm a}=Z_{\rm n}\frac{Z_{\rm c}+jZ_{\rm n}\tan(kw_2)}{Z_{\rm n}+jZ_{\rm c}\tan(kw_2)}+j\Im{(Z_{\rm d})}$.
	Here $Z_{\rm n}=\rho _0c_0/h_2$ and $Z_{\rm c}$ are the acoustic impedance of the neck and the cavity of the Helmholtz resonator, respectively. Im($Z_{\rm d}$) is the radiation impedance which is expressed as: $Z_{\rm d}=\frac{\rho_0c_0}{w_1h_2^2}\frac{1-e^{-jkh_2}-jkh_2}{k^2}+\frac{2k\rho_0c_0}{w_1h_2^2}\sum_{n=1} {\frac{1-e^{-jk_{zn}^\prime h_2}-jk_{zn}^\prime h_2}{{k_{zn}^\prime}^3}}$ with $k_{zn}^\prime=\sqrt[]{k^2-{k_{xn}^\prime}^2}$ and  ${k_{xn}^\prime}=n \pi /w_1$. The acoustic  impedance of the cavity $Z_{\rm c}$ is given by $Z_{\rm c}=\sum_n\rho_0c_0\frac{k(1+e^{2jk''_{xn}w_3})\Phi _n^2}{k''_{xn}h_3(1-e^{2jk''_{xn}w_3})}$, where $\Phi _n=\sqrt[]{2- \delta _n}\cos (n \pi /2) {\rm sinc}(n \pi h_2/2 h_3)$ and $k''_{xn}=\sqrt[]{k^2-(n \pi /h_3)^2}$. The impedance matrix of an arbitrary meta-atom can then be calculated by converting the transfer matrix using 
\begin{equation}
Z=
\begin{bmatrix}
\frac{M_{11}}{M_{21}}     & \frac{M_{11}M_{22}-M_{21}M_{12}}{Z_{21}}  \\
\frac{1}{M_{21}}     & \frac{M_{22}}{M_{21}}
\end{bmatrix}.
\end{equation}

With the theoretical requirement of the impedance matrix profile for perfect wavefront transformation and the versatility of the meta-atom for full control over the bianisotropic response, the next step is to decide the detailed physical dimensions of the meta-atoms that form the metasurface. Since there are three independent elements in the required impedance matrix ($X_{11}$, $X_{12}$, $X_{22}$) and five controlling parameters ($\theta_c$, $w_{\rm a}$, $w_{\rm b}$, $w_{\rm c}$ and $w_{\rm d}$), there can be many combinations for a meta-atom to realize the required impedance matrix. To solve for a practical design within geometrical limitations, a continuous genetic algorithm is adopted for optimization of the design parameters, so that the impedance matrix of the optimized structure matches the theoretical requirements.

We have designed a metasurface to transform a monopole source ($n_1=0$) located at the center to a spinning field with the angular momentum of $n_2=12$. In this case, $r_1=15$~cm, $r_2=20$~cm, and one period is represented by 6 meta-atoms. In this case, each unit cell occupies a sector of $\Delta \phi=\pi/36$, therefore, $S_1=\Delta \phi r_1$ and $S_2=\Delta \phi r_2$. We swept the circumferential positions with a step of 0.1 degrees, and run the GA optimization 50 times at each point to search for the best combination with the lowest relative error.

Although theoretical calculation offers a fast and close approximation of the meta-atom behavior, it will also introduce some error due to truncation of the infinite series and the straight channel assumption. On the other hand, extracting the impedance using commercial simulations (for example, COMSOL Multiphysics) offers slow but more precise characterization. Therefore, based on the structure obtained from theoretical optimization, we further optimize it locally using genetic algorithm by slightly perturbing the structure dimensions within $\pm1$~mm.
The method used for extracting the impedance matrix from simulation was adopted from the standard ``4-microphone" method.
The method uses four microphones to measure the pressure at two fixed points on both sides of the tested structure under two different boundary conditions, and the properties can thus be calculated. 
Based on the same idea, we developed a method to extract the structure properties in cylindrical coordinates. Detailed derivation of the method is summarized in Supplementary Note 3.

\begin{figure}
\includegraphics[width=0.95\linewidth]{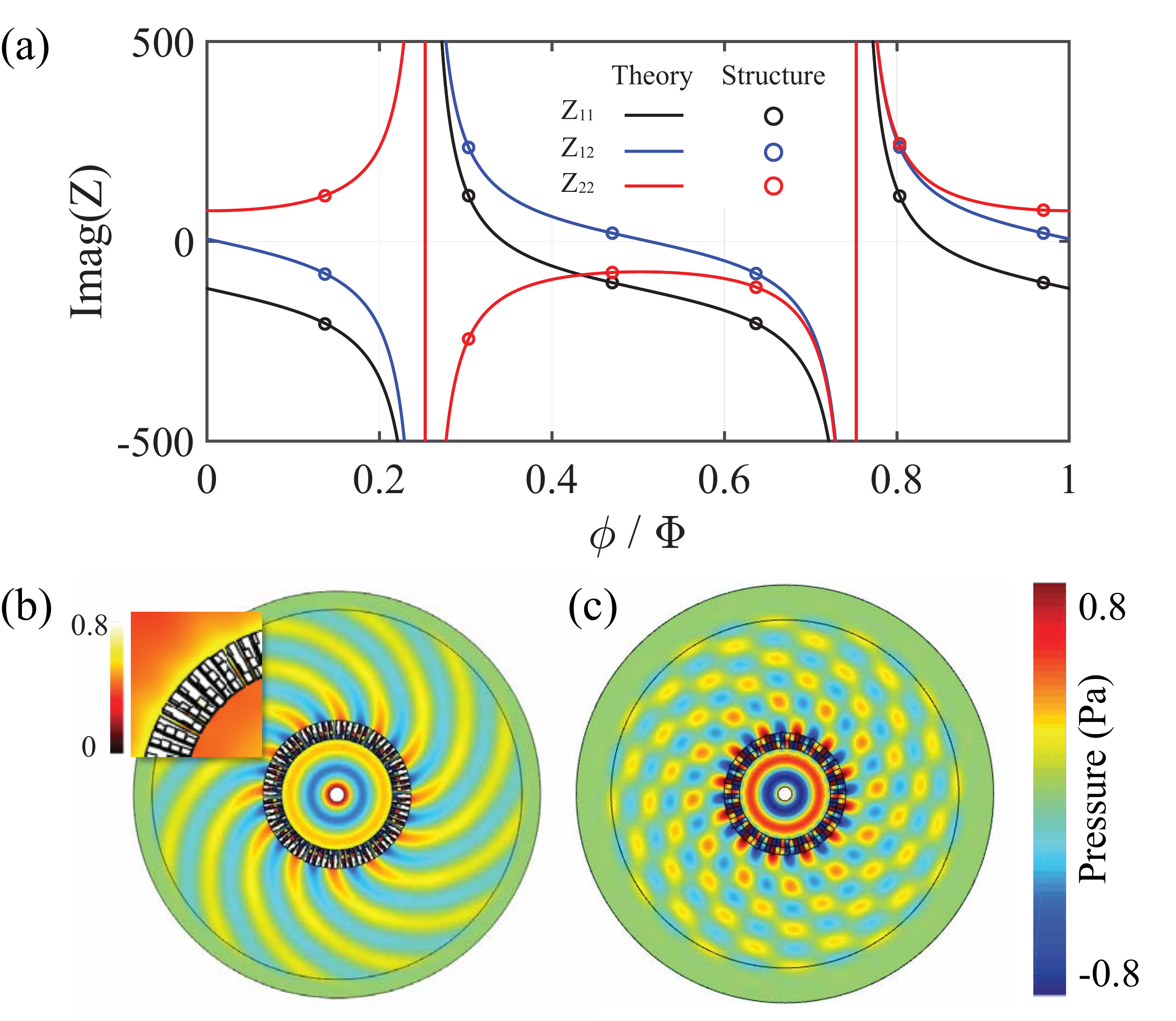}
\caption{Theoretical determined and optimized impedances and the simulated fields. (a) Comparison between theoretical requirements and the achieved values using GA optimization. (b) The real part of the simulated acoustic field using real structures. The inset shows the pressure amplitude near the metasurface. (c) The field generated by GSL based metasurface using ideal unit cells as a comparison.} 
\label{fig:Fig3}\end{figure}

\begin{table}[b]
\caption{\textbf{Design parameters of the meta-atoms}}
\begin{tabular}{ccccccc}
\hline \hline
Cell &  cost$(\%)$ & $\theta_c$ (mm) & $w_a$ (mm) & $w_b$ (mm) & $w_c$ (mm)  & $w_d$ (mm)\\
\hline
1 & 2.08 & 0.5699 & 6.8 & 8.8 & 8.6 & 6.9\\
2 & 0.66 & 0.5655 & 7.0 & 7.0 & 8.3 & 6.4\\
3 & 0.16 & 0.6997 & 8.4 & 7.9 & 2.5 & 5.2\\
4 & 0.55 & 0.7002 & 7.4 & 8.4 & 0.9 & 4.3\\
5 & 0.35 & 1.0221 & 4.1 & 8.1 & 6.7 & 3.0\\
6 & 0.84 & 1.3931 & 8.7 & 2.1 & 0.5 & 3.0\\
\hline \hline
\end{tabular}\label{tab:meta-atoms}
\end{table}

The theoretical requirement for the desired metasurface and the achieved values from the two-step optimization is shown in Fig.~\ref{fig:Fig3}(a). Detailed dimensions of the meta-atoms and their relative errors can be found in Table~\ref{tab:meta-atoms}. We can see that the required impedance is closely realized by the optimized meta-atoms. Simulation of the obtained structure was performed in COMSOL Multiphysics with the pressure acoustics module. The walls of the unit cells are set to be hard due to the large impedance contrast in the implementation. The background medium is air with density 1.21~$\rm{kg/m^{3}}$ and sound speed 343~$\rm{m/s}$. The incident pressure amplitude is 1 Pa at $r=2~\rm{cm}$. The outer edge of the simulated region is connected to a perfectly matched layer. The simulated pressure field and the pressure amplitude are shown in Fig.~\ref{fig:Fig3}(b). We can see that the monopole wavefront is perfectly converted to a field with the angular momentum of 12 without parasitic reflection and scattering. From the pressure amplitude field we can see that the macroscopic transmission coefficient $|T|>1$, i.e., the pressure on the transmission side is larger than the incident side. The corresponding reference GSL metasurface formed by ideal unit cells with the same size and the same number of cells period is shown in Fig.~\ref{fig:Fig3}(c) as a comparison. We can see that there is strong reflection and lots of the transmitted energy is scattered to the unwanted modes and the overall wave pattern is corrupted.

\begin{figure}
\includegraphics[width=1\linewidth]{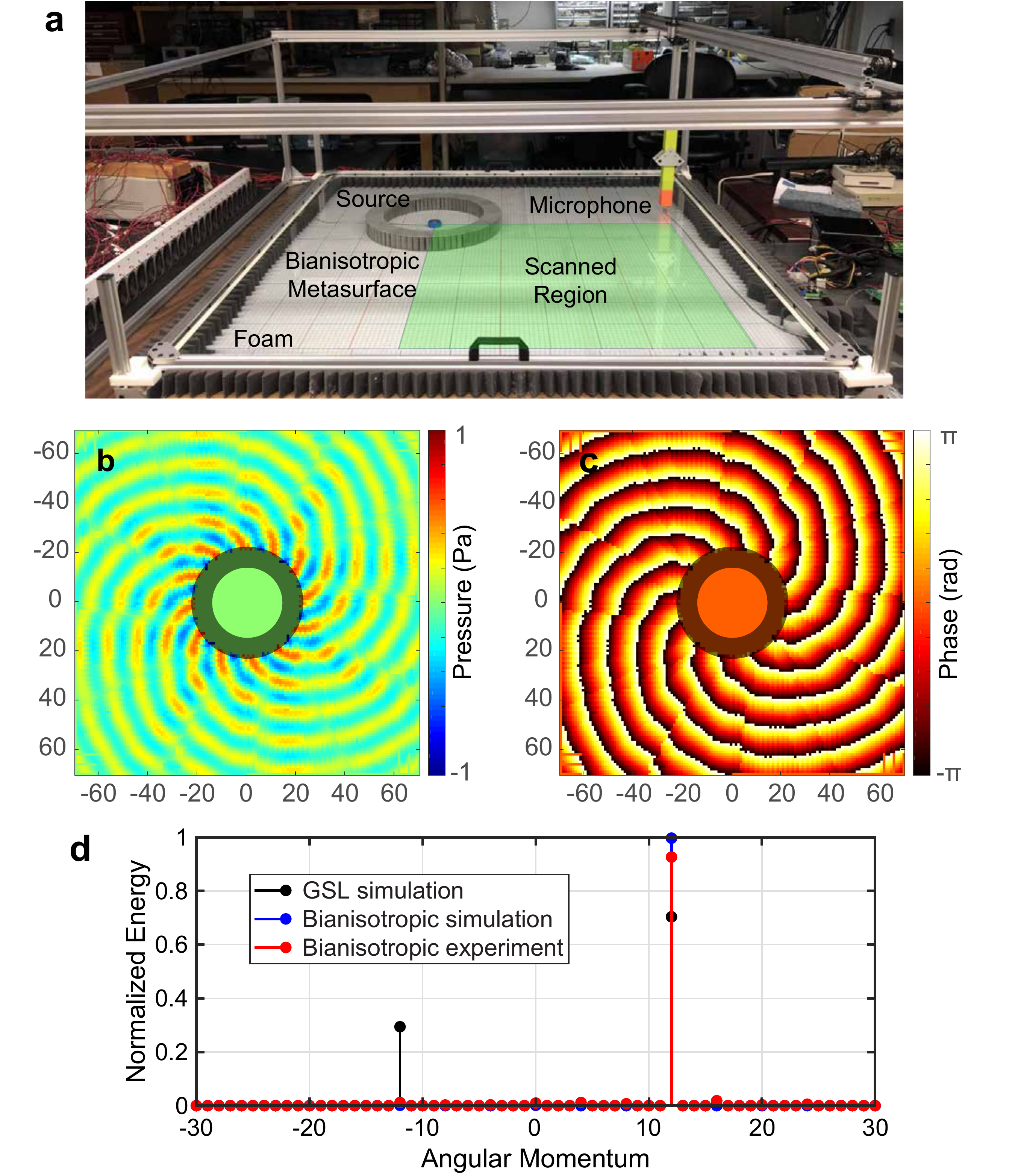}
\caption{Experimental setup and results. (a) A photo of the experimental setup. The field is scanned by moving the microphone in the green region. (b) The real part of the measured pressure field. (c) The phase of the scanned field. We can clearly see that the wavefront is nearly perfectly transformed to the field with the angular momentum $n=12$. (d) The comparison among the bianisotropic metasurface in simulation and experiment, and the ideal GSL-based metasurface in the simulation. In the experiment, 92\% of the transmitted energy is concentrated in the desired mode.} 
\label{fig:Fig6}\end{figure}

\textbf{Experimental Verification.}
The theory and simulations are then verified with experiments. We chose the same example as discussed in the previous section. The experimental setup is shown in Fig.~\ref{fig:Fig6}(a). The sample was fabricated by Selective Laser Sintering (SLS) 3D-printing. The material is Nylon with the density of 950~$\rm{kg/m^{3}}$ and sound speed of 1338~$\rm{m/s}$, so that the walls can be regarded as rigid due to the large impedance contrast with air. The printed sample has the inner radius of 150~mm and the outer radius of 200~mm, and the height of the sample is 41~mm to fit in the 2D-waveguide. The monopole source was provided by a 1-inch speaker located at the center, which sends a Gaussian pulse centered at 3000~Hz. The field was scanned by a moving microphone with a step of 1~cm. Then the field is calculated by performing Fourier transform of the detected pulse. Since the overall size of the scanning system is limited, and the field is symmetric, a quarter of the whole field is scanned, as shown in Fig.~\ref{fig:Fig6}(a), and the measured data is then mapped to other regions. 

The real part of the scanned field and the phase of the field is plotted in Fig~\ref{fig:Fig6}(b) and Fig.~\ref{fig:Fig6}(c). From the experimental results, we can see that the fabricated metasurface created the field with much lower unwanted scatterings compared with an ideal GSL-based metasurface shown in Fig.~\ref{fig:Fig6}(d). The small discrepancy is due to the fabrication errors, and the small difference in the air properties between simulation and experiment. In particular, the sound speed was 344~m/s in our lab during the measurement window, while we assumed 343~m/s in the simulation, which will cause the working frequency to increase by about 8~Hz. To quantitatively characterize the results, we extracted the coefficients of contributing modes by taking the data around a circular trajectory and performing a Fourier transform of the fields at $r=22$~cm to extract the amplitudes of different modes. The power of each mode is calculated and then normalized by the total power. The power distribution over the modes of $n=-30$ to $n=30$ is plotted in Fig.~\ref{fig:Fig6}(d). For comparison, the same analysis is performed for the simulation of the bianisotropic metasurface and the ideal GSL-based metasurface.  We can clearly see that the GSL-based metasurface, even with the perfectly designed cells of full transmission and precise control of the transmitted phase, produces a large component of $n=-12$ mode, so that only 70\% of the transmitted energy is in the desired mode, while in the bianisotropic designs, the unwanted scattering is greatly suppressed, showing 99\% and 92\% of the transmitted energy in the desired mode $n=12$ in simulation and experiment, respectively. The experimental results show good agreement with the simulation, demonstrating the possibility of near perfect transformation of acoustic wavefronts.

\section{DISCUSSION}
In this paper, we have introduced a multi-physics design method for creation of acoustic or electromagnetic bianisotropic metasurfaces of cylindrical shape for perfect generation of waves with arbitrary angular momenta. 
We first defined theoretically the conditions and requirements, and pointed out that controlling the local phase shift in transmission alone cannot achieve such transformations. 
Instead, full control over the bianisotropy is required. 
Then we proposed possible realizations for acoustic waves, and verified them with simulations, showing that the proposed metasurface nearly perfectly transforms a monopole source into a spinning wave field with the angular momentum of 12, which is beyond the ability of conventional GSL-based metasurfaces. Then we proposed a systematic and practical way of creating cylindrical bianisotropic acoustic metasurfaces and verified it with experiments. The experimental results show excellent agreement with simulations, with 92\% of the transmitted energy concentrated in the desired mode, whereas with the use of an ideal GSL-based metasurface, 30\% of the transmitted energy is scattered to other modes. Here we would like to note that the efficiency of the conventional GSL-based design is even lower because the simulation shows that 10\% of the energy is reflected indicating that the ideal efficiency can reach only 63\%, while our design is free of reflections.

The use of waves with non-zero angular momenta has shown great potential in high-speed communications, source illusion and particle manipulation in the fields of optics, electromagnetics, and acoustics. However, one obstacle is the efficiency of generating angular momentum, especially when the target angular momentum is large. In this paper, we have proposed and demonstrated the realization of theoretically perfect generation of angular momenta with a bianisotropic metasurface. We also hope that such metasurfaces can be explored in optics to enhance the efficiency of generating orbital angular momentum beams for high-speed optical communications and other applications.

Here we would like to stress that the proposed design strategy is not only valid for generation of angular momentum beams but for rather general manipulation of wavefronts, both for acoustic and electromagnetic waves. For example, by designing the bianisotropic impedance matrix profile, one may create a multi-polar sources from a single excitation within a limited space; the proposed metasurface may also be applied as an interface between two media to enhance energy transfer; the metasurface may also be applied in topological insulators to either act as a spinning source to excite some certain modes, or even provide the ``pseudo spin'' for topological insulators in airborne systems. We believe that the proposed bianisotropic metasurface concepts can serve as a new approach to designing highly efficient metasurfaces.

\section*{Acknowledgments}
This work was supported by the Multidisciplinary University Research Initiative grant from the Office of Naval Research (N00014-13-1-0631), an Emerging Frontiers in Research and Innovation grant from the National Science Foundation (Grant No. 1641084), and in part by the Academy of Finland (project 287894 and 309421).

\section*{Author contributions}
J.L., A.D. performed theoretical analysis J.L., A.D., and C.S. performed numerical simulations. J.L., C.S. and Z.J. conducted the experiments and processed the data. All authors contributed to analyzing the data and preparing the manuscript. S.A.C. and S.A.T. supervised the study.



\end{document}